\documentstyle[12pt]{article}

\newcommand{\BB}[1]{\mbox{\bf #1}}
\newcommand{\BBn}[2]{\mbox{{\bf #1}$^{#2}$}}

\newcommand{\noi}{\noindent}
\newcommand{\hsp}{\hspace{5.7mm}}
\newcommand{\mini}{\hspace{.3mm}}
\newcommand{\mmini}{\hspace{-.4mm}}
\newcommand{\mimini}{\hspace{-.3mm}}
\newcommand{\mA}{\hspace{-1mm}}

\newcommand{\lsim}{\hspace{1.2mm}\raisebox{.5mm}{$<$}\hspace{-2.9mm}
   \raisebox{-.9mm}{$\scriptstyle\sim$}\hspace{1.7mm}}

\newcommand{\be}{\begin{equation}}
\newcommand{\ee}{\end{equation} \vspace{0mm}}
\newcommand{\ben}{$$}      
\newcommand{\een}{$$ \vspace{0mm}}
\newcommand{\bea}{\begin{eqnarray}}
\newcommand{\eea}{\end{eqnarray} \vspace{0mm}}
\newcommand{\ba}{$\begin{array}{c}}
\newcommand{\ea}{\end{array}$}

\newcommand{\hah}{\hsp\mbox{and}\hsp}

\newcommand{\Qv}{\mbox{\raisebox{-2.2mm}{$\stackrel{\displaystyle
   Q}{\scriptstyle \rightarrow}$}}}
\newcommand{\QT}{\Qv^{\raisebox{-.4mm}{$\scriptstyle\hspace{.3mm}T$}}}
\newcommand{\Qi}{\Qv^{\raisebox{-.4mm}{$\scriptstyle\hspace{.1mm}i$}}}
\newcommand{\QiT}{\Qv^{\raisebox{-.4mm}{$\scriptstyle\hspace{.3mm}
   i\mini T$}}}
\newcommand{\QL}{\Qv_{\raisebox{1.5mm}{$\scriptstyle\hspace{-.2mm}L$}}}

\newcommand{\QR}{\Qv_{\raisebox{1.5mm}{$\scriptstyle\hspace{-.2mm}R$}}}

\newcommand{\QLR}{\Qv_{\raisebox{1.5mm}{$\scriptstyle\hspace{-.2mm}L/R$}}}
\newcommand{\nv}{\mbox{\raisebox{-1.7mm}{$\stackrel{\displaystyle
   n}{\scriptstyle \rightarrow\hspace{.1mm}}$}}}
\newcommand{\mv}{\mbox{\raisebox{-1.7mm}{$\stackrel{\displaystyle
   m}{\scriptstyle \rightarrow\hspace{.1mm}}$}}}
\newcommand{\nvs}{\mbox{\raisebox{-1.5mm}{$\stackrel{\scriptstyle
   n}{\scriptscriptstyle\rightarrow\hspace{.1mm}}$}}}
\newcommand{\Gs}{\Gamma^\ast}
\newcommand{\GLR}{\Gamma\mmini_{L/R}}
\newcommand{\GLRs}{\Gamma\mmini_{L/R}^{\hspace{.4mm}\ast}}
\newcommand{\GL}{\Gamma\mmini_L}
\newcommand{\GR}{\Gamma\mmini_R}
\newcommand{\Qel}{Q_{\mimini\mbox{\scriptsize el}}}
\newcommand{\ov}{\mbox{\raisebox{-1.7mm}{$\stackrel{\displaystyle
   0}{\scriptstyle \rightarrow\hspace{.1mm}}$}}}
\newcommand{\hv}[1]{\mbox{\raisebox{-1.8mm}{$\stackrel{\mini\displaystyle
   h}{\scriptstyle\rightarrow}$}}^{(#1)}}
\newcommand{\vq}[1]{\vec{q}_{#1}}

\newcommand{\Ki}{K^{-1}}
\newcommand{\sH}{\sigma_H}

\newcommand{\ndHt}{\mbox{$n_H/d_H$}}
\newcommand{\eh}{{e^2\over h}}
\newcommand{\eht}{\mbox{$e^2\mmini/\mimini h$}}

\newcommand{\U}{$\mini U(1)$}
\newcommand{\uh}{$\,\hat{u}(1)$}

\newcommand{\bBc}{\mini{\bf B}_c\mini}
\newcommand{\ops}[1]{\varepsilon^{#1}}
\newcommand{\dL}{{\partial \Lambda}}
\newcommand{\Ar}{A|_{\partial \Lambda}}
\newcommand{\da}{\mbox{d}}
\newcommand{\SLas}{S_\Lambda^{\mbox{\scriptsize\hspace{.15mm}as}}}
\newcommand{\BT}{B.T.}     
\newcommand{\WLR}{W\!_{L/R}\mini(\QLR;\mini\Ar)}
\newcommand{\WL}{W\!_L\mini(\QL;\mini\Ar)}
\newcommand{\WR}{W\!_R\mini(\QR;\mini\Ar)}
\newcommand{\GI}{G.I.}     

\newcommand{\halb}{{1 \over 2}}

\newcommand{\PRL}{Phys.\ Rev.\ Lett.\ }
\newcommand{\PRB}{Phys.\ Rev.\ B }

\newcommand{\NPB}{Nucl.\ Phys.\ B }
\newcommand{\AP}{Ann.\ Phys.\ (N.Y.) }
\newcommand{\CMP}{Commun.\ Math.\ Phys.\ }
\newcommand{\SurS}{Surf.\ Sci.\ }

\newcommand{\IJMP}{Int.\ J.\ Mod.\ Phys.\ }

\begin{document}

\begin{flushright}
{\footnotesize\tt K.U.$\mini$Leuven-preprint, May 1994 \\ cond-mat/9406009}
\end{flushright}

\vspace{15mm}

\begin{center}
{\large\bf AN $\mini A\mini D\mini E-{\cal O}\,$
CLASSIFICATION OF MINIMAL\rule[-3.5mm]{0mm}{6mm} \\
INCOMPRESSIBLE QUANTUM HALL FLUIDS$\mini$\footnote{A slightly
modified version of a paper which is to appear in the proceedings
of the NATO

Advanced Research Workshop ``On Three Levels'', Leuven (Belgium),
July 19-24, 1993,

A.~Verbeure {\em et al.}, eds. (Plenum, 1994).}}
\end{center}

\vspace{5mm}

\begin{list}
{}{\setlength{\leftmargin}{25.6mm}\setlength{\parsep}{0cm}
\setlength{\listparindent}{0mm}}
\item
\begin{flushleft}
J\"urg Fr\"ohlich,$^1$ Urban M.~Studer,$^2$ and
Emmanuel Thiran$^1$
\\[5mm]

$^1$Institut f\"ur Theoretische Physik, ETH-H\"onggerberg\\
\hspace{1.6mm}8093 Z\"urich, Switzerland\\
$^2$Instituut voor Theoretische Fysica, K.U.$\,$Leuven\\
\hspace{1.6mm}3001 Leuven, Belgium
\end{flushleft}
\end{list}

\vspace{5mm}

\section{\normalsize\bf INTRODUCTION AND THE GENERAL CLASSIFICATION
PROBLEM}

The quantum Hall (QH) effect$\mini^1$ is observed in
two-dimensional electronic systems (2DES's) subjected to a strong,
uniform, transverse external magnetic field. Experi\-mentally,
such systems are realized as inversion layers that form at the
interfaces of heterostructures (e.g., GaAs/Al$_x$Ga$_{1-x}$As) in
the presence of an electric field (gate voltage) perpendicular to
the structures. To develop an idea of the orders of magnitude
involved in QH systems, we recall that sample sizes are typically
of a few tenths of a $\mini$mm$\mini$ times a few $\mini$mm,
whereas the charge carrier densities, $n = n_{\mbox{\scriptsize
electron}} - n_{\mbox{\scriptsize hole}}$, are of the order of
$\mini 10^{11}$\mmini /cm$^2$, and the magnetic fields, ${\bf
B}_c$, range from about $\mini 0.1\,$T$\mini$ up to $\mini 30\,$T.
Moreover, experiments are performed at very low temperatures, $T$,
typically between $\mini 10\,$mK$\mini$ and $\mini 100\,$mK. An
important quantity characterizing QH systems is the filling
factor $\mini\nu\mini$: denoting by $\,\Phi_{\mimini o}
=h/\mimini e = 4.14 \cdot10^{-11}\,$Tcm$^2$ the magnetic flux
quantum and by $\mini B_{c,\perp}$ the component of the magnetic
field  $\bBc$ perpendicular to a 2DES, the filling factor is
defined by $\,\nu = n\mini\Phi_{\mimini o}/B_{c,\perp}$.

Two basic facts characterize the QH effect. First, the Hall (or
transverse) resistance $R_H$, as a function of $\mini1/\nu$,
exhibits plateaux at heights which are rational multiples of
$\mini h/\mimini e^2$. Second, with the quantization of $R_H$, one
observes a near vanishing of the longitudinal (or magneto-)
resistance $R_L$. [Strictly speaking, this is true only if
measurements are carried out in a {\em stationary$\,$} state and
on {\em large distance scales}, that is, on scales larger
than the phase-coherence length of the constituents of the
systems. For QH systems, the phase-coherence length is of the
order of $\mini 100\mini\mu$m$\mini$ (a so-called ``mesoscopic''
length scale). For a general account of ``quantum interference
fluctuations'' in disordered, mesoscopic systems, see Ref.~2; and,
in particular, for some results in mesoscopic QH systems, see
Ref.~3.] The near vanishing of $R_L$ indicates the {\em absence of
dissipative processes$\,$} in the corresponding QH systems which,
in turn, is interpreted as the presence of an {\em energy
gap$\,$} above the ground-state energy in these many-body systems
(so-called {\em incompressibility}). Incompressibility is a crucial
input to our analysis and, in these notes, we use the term {\em
quantum Hall fluid$\,$} to denote a QH system which exhibits
incompressibility. Next we briefly review the main logical steps
that lead to the general classification problem of QH fluids.

\vspace{5mm}

{\bf Gauge Symmetry and \uh-Current Algebra.} For definiteness,
let us consider a QH fluid which is confined to a domain
$\mini\Omega\mini$ of the $(1,2)$-plane, and let the external
magnetic field, ${\bf B}_c$, be along the 3-axis, i.e., ${\bf B}_c
=(0,0,B_c)$. Furthermore, let us assume that the spins of the
constituent electrons and/or holes are aligned with the strong
field $\mini {\bf B}_c$. [For an analysis of QH fluids including
the spin degrees of freedom, see Refs.~4-6.]

We denote by $\mini Z_\Lambda(A)\mini$ the partition function (at
$T\!=\mmini 0$) of a 2DES confined to a space-time domain
$\,\Lambda=\Omega\times \BB{R}\,$ and coupled to an external
electromagnetic field with associated potential 1-form
$\mini A=\sum_{\mu=0}^2 A_\mu(x)\mini dx^\mu$, where $\mini A_\mu =
A_{\mbox{\scriptsize tot},\mu} - A_{c,\mu}\,$ with $\,\partial_1
A_{c,2} - \partial_2 A_{c,1} = B_c\mini$ fixed. Exploiting the
\U-gauge symmetry of non-relativistic Schr\"odinger quantum
mechanics and the incompressibility $(R_L\mmini=\mimini 0)$ of
QH fluids, one can show$\mini^4$ that, in the limit of large
distance scales and low frequencies ({\em scaling limit$\mini$}),
the {\em effective action$\mini$} of a QH fluid (i.e., the
logarithm of $\,Z_\Lambda(A)$) takes the {\em universal$\,$}
asymptotic form

\be
\SLas(A) \:=\: -{\sH \over 2} \int_\Lambda A \wedge\! \da A\:
+\: \BT(\Ar)\ ,
\label{Seff}
\ee

\noi
where $\,\sH =R_H^{-1}$ is the Hall conductivity of the fluid,
$A \wedge\! \da A = \sum_{\mu\nu\rho} \ops{\mu\nu\rho} A_\mu\mini
\partial_\nu A_\rho\mini$ is the abelian Chern-Simons form, and
$\mini\BT(\Ar)\mini$ denotes a ``boundary term'' depending only on
$\Ar$, the restriction of $\mini A\mini$ to the boundary space-time
$\,\dL$. Since the Chern-Simons term in Eq.~(\ref{Seff}) exhibits
a boundary \U-gauge anomaly, the gauge invariance of the total
effective action, $\SLas(A)$, constrains the form of the boundary
term $\mini\BT(\Ar)$. The cancellation of the gauge anomaly of the
Chern-Simons term can be achieved$\mini^4$ by introducing a set of
$N_L$ and a set of $N_R$ chiral $U(1)$ boundary currents whose
coupling to the external electromagnetic field $\Ar$ is given by
$N_{L/R}$ electric charges which can be arranged in a ``charge
vector'' \mbox{$\QLR$} with $N_{L/R}$ components.
Physically, these left/right chiral boundary systems (CBS's)
describe gapless, chiral electric currents of electrons/holes
circulating at the edge of a QH sample. Denoting by
\mbox{$\,\WLR\mini$} the effective action for these CBS's, the
boundary term $\mini \BT(\Ar)\mini$ is given by

\be
\BT(\Ar)\:=\:\WL - \WR + \GI(\Ar) \ ,
\label{BT}
\ee

\noi
where $\mini\GI(\Ar)\mini$ stands for a possible gauge-invariant
boundary term. Then gauge-anomaly cancellation implies that the
Hall conductivity $\mini\sH\mini$ is given by

\be
\sH \:=\: \Bigl[\mini(\QL\mini,\QL) - (\QR\mini,\QR)\mini \Bigr]\eh
\ , \label{sH}
\ee

\noi
where \mbox{$(\nv\mini,\mv)$} denotes the Euclidean scalar product
of two vectors \mbox{$\mini\nv\mini$} and \mbox{$\mini\mv$}. For
details, see Refs.~4-6; and for related points of view on the role
of edge currents in QH fluids, see the works in Ref.~7.

\vspace{5mm}

{\bf Quantum Hall Lattices.} The vectors \mbox{$\mini\QL\mini$} and
\mbox{$\mini\QR\mini$} are not the only data encoding the physics
of an incompressible QH fluid in the scaling limit. Excitations
above the ground state of the QH fluid are described by unitary
representations of the \uh-current algebras. Among all possible
unitary representations $(\simeq \BBn{R}{N_{L/R}})$ only a subset,
in fact, a lattice, will be realized in a QH fluid. Combining
known properties of these representations$\mini^8$ with the
physical requirement that the spectrum of local excitations of
CBS's be generated by excitations with the quantum numbers of
electrons and holes (i.e., by excitations with electric charges
$\mini\pm e\,$ that obey Fermi statistics and are relatively
local) one infers$\mini^{5,6}$ that the additional data needed to
characterize the scaling limit of a QH fluid is a pair of {\em
odd, positive integral lattices} $\mini\GL$ and $\GR$:
$\GLR$ is generated by $N_{L/R}$ vectors $\mini\vq{(1)},
\ldots,\vq{(N_{L/R})}\mini$ describing distinct one-electron or
one-hole excitations above the ground state. A general vector
$\mini\vec{q}\mini$ in $\GLR$ represents a multi-electron/hole
excitation of the QH fluid. In such a basis, $\GLR$ is fully
specified by its integral Gram matrix $K_{L/R}$

\be
(K_{L/R})_{ij} \:=\: (\vq{(i)},\vq{(j)})\: \in \BB{Z}\,,\ i,j =
1,\ldots,N_{L/R}\ .
\ee

\noi
Fermi statistics of the elementary charge carriers implies the
oddness of the lattice $\GLR$ (i.e., in any basis of
$\GLR$, at least one basis vector has an odd squared
length). The vectors in $\GLR$ do not necessarily label
all physical excitations of the QH fluid. Indeed, any vector in
the dual lattice $\GLRs$ (generated by the vectors
\mbox{$\,\hv{j}\! = \sum_{i=1}^{N_{L/R}}
(\Ki_{L/R})^{ji}\,\vq{(i)}$}, $j=1,\ldots, N_{L/R}\mini$)
describes a (multi-)quasiparticle excitation which is relatively
local to all electrons and holes in the system. However, in
general, such an excitation has fractional electric charge and
anyonic statistics. It is the $N_{L/R}$-dimensional analogue of
the celebrated Laughlin vortices.$^9$

The vector \mbox{$\QLR$} assigns an electric charge to every
boundary excitation. Hence it is a linear functional on the lattice
$\GLR$, i.e., an element of the dual lattice $\GLRs$, and
\mbox{$\,\Qel(\vq{}\,) = \QLR\cdot\vq{}\,$} is the electric charge,
in units of $\,e$, of the excitation labelled by $\mini\vq{}$. Thus
\mbox{$\,\pm 1 = \Qel(\vq{(i)}) = \QLR\cdot\vq{(i)} = (Q_{L/R})_i$},
for every basis vector $\vq{(i)}\in\!\GLR$. This means that
\mbox{$\QLR$} is what mathematicians call a {\em visible}
vector in $\GLRs$. This has an all important consequence:

\ben
(\QLR\mini,\QLR)\;\; \mbox{\em is a rational number.}
\een

\noi
Eq.~(\ref{sH}) then gives the rational quantization of the Hall
conductivity $\mini\sH$, in units of \eht, as observed in QH fluids.

A pair \mbox{$(\Gamma,\mini\Qv)$} is called a {\em quantum Hall
lattice} and represents the geometrical data characterizing a QH
fluid. {\em Classifying universality classes of QH fluids thus
amounts to classifying pairs of QH lattices \mbox{$(\GL,\mini\QL)$}
and \mbox{$\mini(\GR,\mini\QR)$}}.

\vspace{5mm}

\section{\normalsize\bf CLASSIFICATION OF QUANTUM HALL LATTICES}

A lattice $\Gamma$ which is the orthogonal direct sum of
several sublattices is called {\em decompos\-a\-ble} (otherwise
{\em indecomposable}). QH fluids corresponding to decomposable or
indecomposable QH lattices are called {\em composite} or {\em
elementary$\mini$} fluids, respectively. For a composite QH fluid
consisting of two components, we have $\,\sH = \sH^1+\sH^2$, where
\mbox{$\,\sH^i = (\Qi,\Qi)\mini\eht = \Qi\!\cdot \Ki \QiT$\eht},
with \mbox{$\,\ov\neq \Qv^i\in\! \Gamma_{\mmini i}^\ast \subset\Gs,\
i=1,\mini 2\:$}. In the following, we focus our attention on
{\em indecomposable} QH lattices. We note that QH fluids consisting
of electrons $(L)$ {\em and$\,$} holes $(R)$ are composite and thus
will not appear explicitly in the subsequent classification.
Moreover, since the discussions of the two chirality subsectors
($L$ and $R$) are analogous, we may focus on, say, the left one.
We now drop the index $L\mini$ from our notation.

Writing $\,\sH=(\ndHt)\mini\eht$, with $\,\gcd (n_H,\mini d_H)=1$,
Eq.~(\ref{sH}) tells us that, for some positive integer $\,l$,
called the {\em level$\,$} of the QH fluid,

\be
l\mini d_H \:=\: \Delta \:\equiv\: \det K\ , \hah l\mini n_H
\:=\: \gamma \:\equiv\: \Qv \cdot\tilde{K} \QT \ ,
\label{Dg}
\ee

\noi
where $\,\tilde{K}\,$ is the cofactor (or adjoint) matrix of
$\,K$, i.e., $\Ki=\Delta^{-1}\tilde{K}$. Next we state two
general classification results on quantum Hall lattices (for
proofs, see Ref.~6).

First, for a given Hall fraction $\mini\sH\mini$ and two positive
integers $\,l\,$ and $\,N_o$, one can prove that there are
only {\em finitely$\,$} many inequivalent QH lattices
\mbox{$(\Gamma,\mini\Qv)$} with \mbox{$\,\sH=$}
\mbox{$(\Qv\mini,\Qv)\mini\eht =\Qv\cdot \Ki \QT\eht,\;\mini\Delta
= l\mini d_H$}, $\mini$and $\,\dim\Gamma\leq N_o$.

Second, one can show that the {\em minimal, non-trivial fractional
charge}, $e^\ast$, associated with a QH lattice \mbox{$(\Gamma,
\mini\Qv)$} is given by

\be
e^\ast\:\equiv \min_{\nvs\in\mini\mbox{$\scriptstyle \Gamma
\raisebox{1.3mm}{$\scriptscriptstyle\ast$}$}\mimini,\mini
Q_{\mimini\mbox{\tiny el}}(\nvs)\mini \neq\mini 0} |\mini
\Qel(\nv)\mini e \mini | \:=\: {e \over \lambda\mini d_H} \ ,
\label{min}
\ee

\noi
where $\mini\lambda\mini$ is some integer dividing the level $\,l$.
In particular, if $\,d_H\mini$ is {\em even}, one can prove that the
charge parameter $\mini\lambda\mini$ has to be a multiple of $\mini
2\mini$ and the level $\,l\,$ a multiple of $\mini 4$. Thus, for
the observed QH fluids with $\,\sH\, h/\mimini e^2={1\over 2},\mini
({3\over 2})$, and $\mini {5\over 2}$, a {\em
model-independent$\,$} prediction is that $\mini e^\ast\mini$ is a
fraction of $\mini e/4\!:\ e^\ast = (2/\lambda)\, e/4\,$!

Next we focus our attention on {\em minimal$\,$} QH lattices which
are characterized by the property that their levels satisfy
$\,l=1$. There are several basic facts (for proofs, see Ref.~6)
about the corresponding universality classes of {\em minimal
elementary QH fluids$\mini$} which make these classes particularly
attractive from a theoretical, as well as from a
phe\-nom\-eno\-log\-i\-cal, point of view:
\rule[-4mm]{0mm}{5mm}

{\bf (M1) }For the fractions $\,\sH = 1/(2\mini p+1)\,\eht,\
p= 1,2,\ldots\,$, one can easily show that the famous Laughlin
fluids$\mini^9$ do correspond to the {\em unique} minimal
indecomposable QH lattices \mbox{$(\Gamma,\mini\Qv)$} where:
$\Gamma$ is one-dimensional, it is generated by a vector
$\mini\vq{(1)}\mini$ with length squared $\,(\vq{(1)},\vq{(1)}) =
2\mini p+1 = K_{11} = K$, the visible vector \mbox{$\mini\Qv\mini$}
is given by \mbox{$\mini\Qv = \hv{1}\mini$} where \mbox{$\,\hv{1}\!
= 1/(2\mini p+1)\,\vq{(1)}\,$} is generating the dual lattice $\Gs$,
and $\,\sH\,h/\mimini e^2 = (\Qv,\Qv) = 1/(2\mini p+1)$.
\rule[-4mm]{0mm}{5mm}

{\bf (M2) }If $\,l=1\,$ then $\,d_H=\det K\,$ has to be {\em
odd$\mini$}. Experimentally, there is ample evidence for the
``odd-denominator rule''.$^{10}$ For {\em one}-layer (or
{\em one}-component) systems, the only firmly established exception
to this rule is a QH fluid with $\,\sH={5\over 2}\mini\eht$.$^{11}$
[For a discussion of even-denominator QH fluids, see Refs.~5
and~6.] Furthermore, the following relationship holds between the
numerator $\,n_H\,$ of the Hall conductivity $\mini\sH\mini$ and
the dimension $\,N= \dim\Gamma\,$ of the indecomposable QH lattice:

\be
\mbox{for }\, n_H\;\left\{ \begin{array}{c} \!\mbox{\em even} \\
\!\mbox{\em odd} \end{array} \right\} \, ,\ N\, \mbox{ has to be }
\left\{ \begin{array}{l} \!\mbox{\em even} \\ \!\mbox{{\em
odd\hspace{.1mm}},$\,$ and }\, N\equiv n_H\mA \pmod{4} \end{array}
\right\} \ .
\label{dim}
\ee

{\bf (M3) }For minimal QH fluids, one can prove a {\em
charge-statistics theorem$\mini$}: for such fluids, the (anyonic)
statistical phases, \mbox{$\theta(\nv) = (\nv\mini,\nv) =
\nv\mmini\cdot\Ki \nv^T\mmini$}, of (quasiparticle) excitations,
labelled by \mbox{$\nv \in\!\Gs$}, are completely
determined by the single quantity of their (fractional) charges,
\mbox{$\Qel(\nv) = (\Qv\mini,\nv) = \Qv\cdot\Ki\nv^T$}. [In
non-minimal QH fluids, there are anyons which have the same
fractional charges but different statistical phases!]
Experimentally, there is evidence$\mini^{12}$ that the observed
odd-denominator QH fluids exhibit ``elementary'' anyons with
fractional charges given by $\,e^\ast = e/d_H$. Eq.~(\ref{min})
then suggests that $\,\lambda =1$, which is a property
automatically satisfied by minimal QH fluids where $\,\lambda = l=
1\mini$!
\rule[-4mm]{0mm}{5mm}

Conway, Sloane, and Sloane$\mini^{13}$ have compiled a
classification of all positive integral quadratic forms $K\mini$
associated with an indecomposable (even or odd) lattice $\Gamma$ for
a range of determinants given by $\,1\leq \Delta\equiv\det K \leq
25\,$ and for lattice dimensions up to a limit ranging from $18$
(for $\,\Delta = 1$) to $7$ (for $\,\Delta = 25$). This
classification involves over $100$ odd lattices of odd determinant.
When combined with a systematic study of visible vectors
\mbox{$\mini\Qv\mini$} in the associated dual lattices $\Gs\mimini$,
it yields, for a physically interesting range of Hall conductivities
$\mini\sH$, {\em all$\,$} possible {\em minimal indecomposable QH
lattices} \mbox{$(\Gamma,\mini \Qv)$} of dimension $\mini N=
\dim\Gamma\leq N_\ast(\sH)$, where the dimensions $\mini
N_\ast(\sH)\mini$ depend on the limits in Ref.~13 and
satisfy~(\ref{dim}). We summarize our results in Table~1 below,
where the dimensions $\mini N_\ast(\sH)\mini$ are indicated in
square brackets to the right of each fraction $\mini\sH$, except for
fractions with $\,n_H=1\,$ for which the Laughlin fluids are unique
(see~{\bf (M1)}).

In Table~1, minimal indecomposable QH lattices (or, equivalently,
universality classes of minimal elementary QH fluids) are
specified in terms of explicit matrix and vector realizations of
the pairs \mbox{$(K,\mini \Qv)$}. [For a general discussion of the
geometry of QH lattices making use of gluing theory, see Ref.~6.]
The results are presented relative to lattice and dual lattice
bases in $\Gamma$ and $\Gs\mmini$, respectively, which make
the {\em symmetries$\mini$} of the corresponding QH fluids most
manifest.$\mini^{4-6}$ The general results of Refs.~13 and 6 show
that most QH lattices in the range of determinants and dimensions
specified above have a  Gram matrix of the form

\be
K\:=\,
\renewcommand{\arraystretch}{1.8}
\left( \begin{tabular}{c|c|c}
$\cal O$ & \renewcommand{\arraystretch}{.8}\ba \underline{t}_1
\\ \underline{s}_1\ea &
\renewcommand{\arraystretch}{.8}\ba
\underline{t}_2 \\ \underline{s}_2\ea\rule[-4mm]{0mm}{6mm}  \\
\hline
$\underline{t}^T_1\ \underline{s}^T_1$ &
$C(X_1)$\rule[-4mm]{0mm}{6mm} & 0  \\ \hline
$\underline{t}^T_2\ \underline{s}^T_2$ & 0 & $C(X_2)$
\end{tabular} \right)\, ,
\label{nota}
\ee

\noi
which we abbreviate by writing ${\{ {\cal O}\mini |\mini
\mbox{}^{t_1,\mini s_1}\!X_1,\mini \mbox{}^{t_2,\mini s_2}\!X_2
\}}_N$, where $\,C(X_i)\,$ denotes the {\em Cartan matrix$\,$} of a
simply-laced simple Lie algebra, $X_i$, with non-trivial
center, $i=1,2$. For $\mini i=1,2,\ X_i\mini$ is one of the
following algebras: $A_{n-1}=su(n),\ n=2,3,\ldots\,,\ D_n=so(2n),\
n=4,5,\ldots\,,\ E_6$, or $\,E_7$. [$\,E_8$-sublattices do not
appear in elementary QH fluids! For the Cartan matrices we use the
explicit forms given, e.g., in Ref.~14.] Furthermore,
in~(\ref{nota}), $\cal O\,$ stands either for an odd integer,
${\cal O} = 2\mini p+1,\ p=1,2,\ldots\,$, or for an odd,
two-dimensional integral quadratic form with Gram matrix
\mbox{\renewcommand{\arraystretch}{.7}$ {\cal O}= \left(
\begin{array}{cc} \mA a & \mA b\mA \\ \mA b & \mA c \mA\end{array}
\right)$}, abbreviated by $\,a\mini^bc\,$; $\,N= \mbox{rank}\mini
K=\dim\Gamma$. The quantities $\mini\underline{t}_{\mini i} \mini$
and $\mini\underline{s}_{\mini i}\mini$ denote (rank$\mini
X_i$)-dimensional integral vectors of the form
$\,\underline{t}_{\mini i} = (0,\ldots, 0,-1, 0,\ldots,0)$, where
$\,-1\,$ stands in the $\,t_i^{\mini\mbox{\scriptsize th}}$ place
indicated by $\,\mbox{}^{t_i,\mini s_i}\!X_i$, and similarly for
$\mini\underline{s}_{\mini i},\ i=1,2$. [The vectors
$\mini\underline{s}_1 \mini$ and $\mini\underline{s}_2\mini$ are
understood to be absent if $\,\dim {\cal O}=1$.] Finally, if
$\,t_i$ or $\,s_i=0\,$ then $\mini\underline{t}_{\mini i}\mini$ or
$\mini\underline{s}_{\mini i}=\underline{0}\mini$.

A particularly interesting subclass of minimal indecomposable
QH lattices is formed by what we call {\em maximally symmetric$\,$}
lattices. A maximally symmetric QH lattice has the property that
the neutral sublattice $\Sigma$ of $\mini\Gamma$ (which
lies in the orthogonal complement of \mbox{$\,\Qv\mini$}, i.e.,
\mbox{$\Qel(\nv) = \Qv\cdot\Ki\nv^T = 0$}, for
\mbox{$\nv\in\!\Sigma\,$}) is a direct sum of $A_n,\mini D_n$, and
$E_{n\neq 8}$-lattices, and nothing else. In particular, they have
$\,\dim {\cal O}=1\,$ and, relative to the dual bases chosen
in~(\ref{nota}), their visible vectors are always of the form
\mbox{$\,\Qv = (1,0,\ldots,0)$}. One can show$\mini^{4-6}$ that
the corresponding maximally symmetric QH fluids exhibit {\em
Kac-Moody algebras$\,$} at level $1$ associated with the Lie
algebras $X_1$ and $X_2$.

In Table~1, we use round brackets ${( {\cal O}\mini |\mini
\mbox{}^{t_1,\mini s_1}\!X_1,\mini \mbox{}^{t_2,\mini s_2}\!X_2 )}_N$
to indicate maximally symmetric QH lattices, thereby distinguishing
them from all other possible minimal in\-de\-com\-pos\-a\-ble QH
lattices (with $\,N \leq N_\ast(\sH)$) for which we use square
brackets ${[ \ldots ]}_N$. Moreover, for maximally symmetric QH
lattices, we do not display the vectors \mbox{$\mini\Qv\mini$}
explicitly.

It seems natural to call maximally symmetric QH fluids
``{\em generalized Laughlin fluids$\,$}'' because they can be
constructed from Laughlin's $1/(2\mini p+1)$-fluids by ``adding
symmetries and iterating''. Let

\be
K \:=\, \left( \begin{tabular}{c|c}
\raisebox{-2.5mm}{$K'$} & $\hspace{-1mm}\underline{t}$
\\  \cline{2-2}
 & $0$\rule[0mm]{0mm}{4.3mm}
\\ \hline
\begin{tabular}{c|c}
$\underline{t}^T$ & $\ 0$\rule[-3mm]{0mm}{8mm}
\end{tabular}
& $C(X)$
\end{tabular} \right)\ , \hah
\Qv
\:=\,(\mini\overbrace{\underbrace{1,0,\ldots,0}_{\mbox{\scriptsize
rank$\mini K'$}}}^{\Qv'},\underbrace{0, \ldots,
0}_{\mbox{\scriptsize rank$\mini X$}}\,) \ ,
\ee

\noi
similarly to~(\ref{nota}). Then one easily proves the following
relations for
\rule[-6.6mm]{0mm}{8mm}

\renewcommand{\arraystretch}{1.3}
\begin{tabular}{lclcl}
\hspace{-9mm}$\mbox{}^t\!X=\mbox{}^t\mA
A_{n-1}=\mbox{}^t\mmini su(n)$ & $\mA:\mA$ & $\gamma = n\mini
\gamma'$ & and & $\Delta = n\mini \Delta' - t(n-t)\mini
\gamma',\ $ for $\,t=1,\ldots,\mini n-1\,$, \\
\hspace{-9mm}$\mbox{}^t\!X=\mbox{}^t\!D_n=\mbox{}^t\mmini
so(2n)$ & $\mA:\mA$ & $\gamma = 4\mini \gamma'$ & and &
$\Delta = 4\mini \Delta' - n\mini\gamma',\;\:$ if $\:t=
n-1\mini$ or $\,n\,$, \\
\hspace{-9mm}$\mbox{}^t\!X=\mbox{}^t\!E_6$ & $\mA:\mA$ &
$\gamma = 3\mini\gamma'$ & and & $\Delta = 3\mini \Delta' -
4\mini\gamma',\;\:$ if $\:t=1\mini$ or $\,5\,$, \\
\hspace{-9mm}$\mbox{}^t\!X=\mbox{}^t\!E_7$ & $\mA:\mA$ &
$\gamma = 2\mini\gamma'$ & and & $\Delta = 2\mini \Delta' -
3\mini\gamma',\;\:$ if $\:t=6\,$,
\end{tabular}
\vspace{-9.7mm}
\be
\mbox{ }
\label{symit}
\ee
\vspace{-1.9mm}

\noi
where $\,\sH'=(\gamma\mini'/\Delta')\mini\eht$, with $\,\Delta' =
\det K'\,$ and \mbox{$\,\gamma\mini'= \Qv'\mmini\cdot\tilde{K}'
\Qv'^{\mini T}\mmini$}, and accordingly for the unprimed
quantities; see also~(\ref{Dg}).

We note that, for a given value of $\mini\sH$, one can construct
all maximally symmetric QH lattices, in arbitrary dimension,
reproducing that value of $\mini\sH$. In Table~1, {\em all$\,$}
minimal, maximally symmetric QH lattices are given for the
physically relevant range of Hall conductivities $\,1/7\leq \sH\,
h/\mimini e^2 = \ndHt \leq 3\,$ with $\,n_HÊ\leq 12\,$ and
$\,d_H\leq 17$, and all QH lattices that can be constructed on the
basis of the classification in Ref.~13 are displayed. For
additional results, see Refs.~5 and~6.

A step towards a comparison of our results with phenomenological
data obtained in one-layer (or one-component) QH
systems$\mini^{10}$ is made, in Table~1, by indicating in bold
type all experimentally observed fractions. Moreover, Hall
fractions for which there is some evidence are typed in bold and
enclosed in brackets. All other fractions (plain) have so far not
been established experimentally in one-layer systems. A
discussion of our results can naturally be organized by collecting
QH lattices of the same ``{\em symmetry type$\,$}'' into ``{\em
structural families$\,$}''; see Refs.~5 and~6. Moreover, in these
references, detailed discussions of physical implications of our
results above and of further complete investigations of some more
general classes of QH lattices than the minimal ones are given, as
well as comparisons with the standard Haldane-Halperin$\mini^{15}$
and Jain-Goldman$\mini^{16}$ hierarchy schemes. Here we only
mention that Table~1 comprises the Laughlin $\mini1/(2\mini
p+1)$-fluids, the ``basic'' Jain fluids$\mini^{17}$ (corresponding
to the $A$-type fluids with $\,\sH < \halb\mini\eht\:$!), and the two
two-dimensional hierarchy fluids with $\mini\sH={4\over
11}\mini\eht$ and $\mini{6\over 17}\mini\eht$. All other
(universality classes of) QH fluids given in Table~1 do not seem to
have appeared previously in the literature. Evidence that some of
these ``new'' QH fluids might actually have been observed in some
experiments will appear in Refs.~5 and 6.

\vspace{5mm}

\section{\normalsize\bf ACKNOWLEDGMENT}

UMS would like to thank the Onderzoeksfonds K.U.~Leuven (grant
OT/92/9) for financial support.

\newpage

\vspace*{50mm}

{\bf Table 1. } Minimal indecomposable quantum Hall lattices with
$\mini 1/7 \leq \ndHt =\sigma_H\,h/\mimini e^2 \leq 3$.
$\mini{(\ldots)}_N\mini$ and $\mini\Qv =(1,0,\ldots,0)$: maximally
symmetric lattices (complete list); $\mini{[ \ldots]}_N\mini$ and
$\mini\Qv\mini$ as indicated: non-maximally symmetric lattices (all
with $N\leq N_\ast(\sH)$).

\vspace{15mm}

{\tt [The LaTeX file for Table~1 is appended to the text and uses
the

config file A4 Landscape.]}

\newpage

\section{\normalsize\bf REFERENCES}

{\small
\noi
\mbox{$\hspace{2mm}$}1. K.~von Klitzing, G.~Dorda, and M.~Pepper,
\PRL {\bf 45}, 494 (1980);

D.C.~Tsui, H.L.~Stormer, and A.C.~Gossard, \PRB {\bf 48}, 1559
(1982);

for a review, see, e.g., R.E.~Prange and S.M.~Gervin, eds., {\em
The Quantum Hall Effect},

Second Edition, Graduate Texts in Contemporary Physics (Springer,
New York, 1990).

\noi
\mbox{$\hspace{2mm}$}2. B.L.~Al'tshuler and P.A.~Lee, Physics
Today {\bf 41} (12), 36 (1988);

R.A.~Webb and S.~Washburn, {\em ibid.}~{\bf 41} (12), 46 (1988).

\noi
\mbox{$\hspace{2mm}$}3. R.~Mottahedeh {\em et al.}, Solid State
Commun. {\bf 72}, 1065 (1989);

D.~Yoshioka, J.\ Phys.\ Soc.\ Jpn. {\bf 62}, 839 (1993).

\noi
\mbox{$\hspace{2mm}$}4. J.~Fr\"ohlich and U.M.~Studer, \CMP {\bf
148}, 553 (1992); Rev.\ Mod.\ Phys.

{\bf 65}, 733 (1993).

\noi
\mbox{$\hspace{2mm}$}5. J.~Fr\"ohlich, U.M.~Studer, and E.~Thiran,
``Structuring the set of incompressible

quantum Hall fluids'', in preparation; ``Classification of quantum Hall
fluids'', in

preparation.

\noi
\mbox{$\hspace{2mm}$}6. J.~Fr\"ohlich and E.~Thiran, ``Integral quadratic
forms, Kac-Moody algebras, and fractional

quantum Hall effect: an $ADE-O\mini$ classification'', preprint,
ETH-TH/93-22, to appear

in J.~of Stat.~Phys. (July 1994).

\noi
\mbox{$\hspace{2mm}$}7. B.I.~Halperin, \PRB {\bf 25}, 2185 (1982);

M.~B\"uttiker, {\em ibid.}~{\bf 38}, 9375 (1988);

C.W.J.~Beenakker, \PRL {\bf 64}, 216 (1990);

A.H.~MacDonald, {\em ibid.}~{\bf 64}, 220 (1990);

X.G.~Wen, {\em ibid.}~{\bf 64}, 2206 (1990); \PRB {\bf 41}, 12838
(1990);

J.~Fr\"ohlich and T.~Kerler, \NPB{\bf 354}, 369 (1991);

M.~Stone, \AP {\bf 207}, 38 (1991);

R.C.~Ashoori {\em et al.}, \PRB {\bf 45}, 3894 (1992);

K.~von Klitzing, Physica {\bf B 184}, 1 (1993).

\noi
\mbox{$\hspace{2mm}$}8. P.~Goddard and D.~Olive, \IJMP A {\bf 1},
303 (1986).

\noi
\mbox{$\hspace{2mm}$}9. R.B.~Laughlin, \PRL {\bf 50}, 1395 (1983); \PRB
{\bf 27}, 3383 (1983).

\noi
10. D.C.~Tsui, Physica B {\bf 164}, 59 (1990), and references
therein;

T.~Sajoto {\em et al.}, \PRB {\bf 41}, 8449 (1990), and references
therein;

H.W.~Jiang {\em et al.}, \PRB {\bf 44}, 8107 (1991);

H.L.~Stormer, Physica B {\bf 177}, 401 (1992), and references
therein.

\noi
11. R.L.~Willett {\em et al.}, \PRL {\bf 59}, 1776 (1987);

J.P.~Eisenstein {\em et al.}, {\em ibid.}~{\bf 61}, 997 (1988); \SurS {\bf
229}, 31 (1990).

\noi
12. R.G.~Clark {\em et al.}, \PRL {\bf 60}, 1747 (1988);

S.W.~Hwang {\em et al.}, \SurS {\bf 263}, 72 (1992).

\noi
13. J.H.~Conway, F.R.S.~Sloane, and N.J.A.~Sloane, Proc.\ R.\
Soc.\ Lond.\ A {\bf 418}, 17 (1988),

and references therein.

\noi
14. R.~Slansky, Phys.~Reports {\bf 79}, 1 (1981).

\noi
15. F.D.M.~Haldane, \PRL {\bf 51}, 605 (1983);

B.I.~Halperin, {\em ibid.}~{\bf 52}, 1583 (1984).

\noi
16. J.K.~Jain and V.J.~Goldman, \PRB {\bf 45}, 1255 (1992).

\noi
17. J.K.~Jain, \PRL {\bf 63}, 199 (1989); \PRB {\bf 41}, 7653
(1990); Comments

Cond.~Mat.~Phys.\ {\bf 16}, 307 (1993).
}

\end{document}